\documentclass[prb,graphicx,preprint,showpacs,superscriptaddress]{revtex4-1}
\usepackage{amsmath,amssymb,amsthm,color}
\usepackage{graphicx}
\usepackage[english]{babel}
\usepackage{bm}

\begin{document}


\title{Evolution and dimensional crossover from the bulk subbands in ABC-stacked graphene to a three-dimensional Dirac cone structure in rhombohedral graphite} 



\author{Ching-Hong Ho}
\affiliation{Center for General Education, Tainan University of Technology, 710 Tainan, Taiwan}

\author{Cheng-Peng Chang}
\email[]{t00252@mail.tut.edu.tw}
\affiliation{Center for General Education, Tainan University of Technology, 710 Tainan, Taiwan}

\author{Ming-Fa Lin}
\affiliation{Department of Physics, National Cheng Kung University, 701 Tainan, Taiwan}

\date{\today}

\begin{abstract}
Rhombohedral graphite behaves as a topological semimetal, possessing flat surface subbands while being semimetallic in the bulk. The bulk-surface correspondence arises from the ABC-stacking configuration of graphene layers. The bulk subbands in rhombohedral graphite can be interpreted as a three-dimensional Dirac cone structure, whose Dirac points form continuous lines spiraling in momentum space. In this paper, we studied the evolution of the gapped bulk subbands in ABC-stacked $N$-layer graphene in the increase of $N$ and their dimensional crossover to the $3$D Dirac cone structure in the bulk limit, where the bulk gap closes up at the Dirac-point spirals. In order to clarify the effect of coupling to the surface subbands, we use a non-perturbative effective Hamiltonian closed in the bulk subspace. As a consequence, the wave length of the standing wave function across the stack of layers depends on the in-plane Bloch momentum. In the bulk limit, the coupling vanishes and hence the wave length is irrelevant to the surface.

\end{abstract}

\pacs{71.20.-b, 73.21.Ac, 73.22.Pr}

\maketitle 

\par\noindent
\section{INTRODUCTION}
The research on graphene layers and their stacks has lasted till today. The inherent symmetrical and topological properties inspire new ideas in many fields of physics. According to the hexagonal lattice symmetry, a number of graphene layers can be so stacked that any adjacent two are shifted from each other in either direction along the armchair orientation. Among others, ABC-stacking configuration is especial in that all the layers are shifted in a single direction. The non-trivial topological phase thus induced can give rise to surface states,\cite{xiao11,heikkila11} which are localized at the outermost layers and coupled to the rest, so-called bulk states.\cite{xiao11,heikkila11,guinea06,otani10} By contrast, AB-stacking configuration does not accommodate any surface states because of the relevant trivial topology.\cite{xiao11} As having been realized in ABC-stacked $N$-layer graphene (ABCNG),\cite{kumar13} the surface states are characterized by flat subbands crossing about the zero energy. The related chiral effective Hamiltonian with chirality $J=N$ was constructed,\cite{min08,koshino09,zhangf10} and associated quantum Hall effect (QHE) were predicted and observed at least for $N=3$.\cite{min08,kumar11,yuan11,zhangl11}

Apart from the surface subbands, the bulk subbands are gapped and the surface subbands reside in the bulk gap.\cite{xiao11,heikkila11,zhangf10} The band structure can be changed by expermimental means of a substrate or gate, which induce on-site Coulomb potential difference between graphene layers. The major change is the splitting of the crossing surface subbands,\cite{ohta07,koshino10} as a consequence of the broken inversion symmetry in ABC-stacking configuration. Remarkably, as the thickness $N$ increases the splitting diminishes and vanishes ultimately, implying a topological robust of the surface subband in the bulk limit. For the observation of QHEs, the ideal condition without Coulomb potential is also experimentally feasible.\cite{ohta07} The surface subbands are thus more and more flat with increasing $N$ while the bulk subbands evolve to be even more complex, as theoretically known.\cite{xiao11,koshino10} So far, the bulk subbands have been seldom studied, probably because no attractive properties are expected from them. Their crossover in the three-dimensional ($3$D) bulk limit is, however, noticed in recent researches.\cite{xiao11,heikkila11}

In the bulk limit, the bulk lattice of ABCNG comprises enormously many graphene layers and usually named rhombohedral graphite (RG). There is a crossover of the lattice symmetry from being hexagonal for finite $N$ to being rhombohedral for infinite $N$. The rhombohedral lattice has a two-atom primitive unit cell and the $2\times2$ Hamiltonian is also chiral.\cite{xiao11} The derived bulk subbands are gapless, and can be interpreted as a $3$D Dirac cone structure with continuous locations of Dirac points (DPs).\cite{ho13,ho14} The lines of DPs spiral in momentum space in association with a sausage-like Fermi surface.\cite{ho13,mcclure69} In consistence, a previous semi-infinite analysis of ABCNG has shown that the resulting density of states (DOS) and Landau level spectrum are the same as in monolayer graphene.\cite{guinea06} The $3$D Dirac cone structure in RG can explain the $3$D QHE observed in graphite samples,\cite{arovas08} where $2$D Dirac fermions of chirality $J=1$ are transported in each layer and the quantized conductivities are the same as in monolayer graphene.\cite{kempa06} This $2$D transport character is a general attribute of graphite, for which the much smaller ratio of the interlayer to the intralayer hopping is responsible. In principle, the effective dimensionality should be determined by the interplay between the interlayer hopping and the interlayer electron-electron interaction.\cite{lopezsancho07} The latter does not affect the characteristics of the low-energy band structure obtained from single-particle models. In that interplay, however, the $3$D nature of RG shows up as the presence of optical magnetoplasmons, which is absent in monolayer graphene.\cite{ho15}

As a whole, RG has a bulk-surface correspondence owing to the non-trivial topology,\cite{xiao11,heikkila11} just as the bulk-edge correspondence in monolayer graphene with zigzag edges.\cite{hatsugai09} This correspondence provides a topological reasoning for why the splitting of the surface subbands of ABCNG vanishes in the bulk limit, even in the presence of on-site Coulomb potential.\cite{ohta07} The surface subbands are protected by the DP spirals, which behave as nodal lines similar to the DPs as nodal points in the bulk of monolayer graphene.\cite{heikkila11,burkov11} Consequently, RG possesses robust flat surface subbands while being semimetallic in the bulk with a $3$D Dirac cone structure. Therefore, a comparison might be made to certain ABC-layered topological insulators, e.g., Bi$_{2}$Se$_{3}$ and Bi$_{2}$Te$_{3}$, \cite{zhangh09} for which the dimensional crossover from $3$D to $2$D has been observed.\cite{zhangy10}

It is of interest to understand the band structure of ABCNG ranging from $2$D to $3$D, as a comparable study to those on AB-stacked $N$-layer graphene.\cite{koshino10,partoens07,koshino07} Here we focus on the bulk subbands. In previous first-principle calculations the bulk gap was known to be still open for $N=10$,\cite{xiao11} and even in analyses for arbitrary $N$ the dimensional crossover was not elucidated.\cite{koshino10} Furthermore, these works were conducted by simultaneously considering the surface and the bulk subbands so that the effect of coupling between them is not easy to clarify. Our purpose is to explore the evolution of the gapped bulk subbands under the coupling effect in the increase of $N$ and to show the dimensional crossover to the gapless $3$D Dirac cone structure in the bulk limit. This paper is organized as follows. In Sec. \uppercase\expandafter{\romannumeral 2} we briefly describe the $3$D Dirac cone structure in the continuous approach. Then we set forth the chain model for ABCNG, with regard to its mapping relation to the continuous approach. It is shown that the infinite chain model leads to the $3$D Dirac cone structure as well. Based the finite chain model, we construct non-perturbatively an effective Hamiltonian that is closed in the bulk subspace, so as to embody the coupling effect. In Sec. \uppercase\expandafter{\romannumeral 3}, we resolve the second-order recursion involved in the eigenproblem of the effective bulk Hamiltonian and solve out the eigenenergies and eigenmodes for arbitrary $N$. The eigenmodes are characterized so that the indices for the bulk subbands are related to the bulk wave numbers. Moreover, we calculate the associated bulk DOS. In Sec. \uppercase\expandafter{\romannumeral 4}, we analyze the obtained bulk subbands in ABCNG to elucidate their evolution and dimensional crossover. A summary with an outlook is given in Sec. \uppercase\expandafter{\romannumeral 5}.

\par\noindent
\section{MODEL}
The stacking configuration of ABCNG is shown in Fig. 1(a), where graphene layers infinitely extended in the $(x,y)$ plane and stacked along the $z$ direction are labeled by $l(=1,2,...,N)$. A solid view of the lattice of ABCNG is given in Fig. 1(b), with the lattice constant of a single layer $a=0.246$ nm and interlayer distance $d=0.337$ nm. Carbon atoms in this configuration are classified into to two sets of atomic sublattices. One set contains the surface sublattices ($B_{1}$ and $A_{N}$); the other set contains the bulk sublattices. Each bulk atom is vertically bonded with another one that is sited in either the adjacent upper or the adjacent lower layer. The surface atoms in the two outermost layers are free from such bonding. If $N$ goes infinity, Fig. 1(b) also shows the bulk lattice of RG. In the bulk of RG, all $A_{l}$ ($B_{l}$) are indistinguishable, denoted by $A$ ($B$) and hence define a two-atom rhombohedral primitive unit cell and an alternative hexagonal non-primitive unit cell of triple volume. The bulk lattice of RG is spanned by the primitive unit vectors $\mathbf{a}_{1,2,3}$, which add up to the $c$-axis ($(111)$) pointing in the $z$ direction. In Fig. 2(a), the first rhombohedral Brillouin zone (BZ) is schematically depicted, accompanied with a prism of the folded hexagonal zone of the same height $\pi/d$. As noticed, the vertical edge lines of the folded hexagonal zone, through the $K^{(\xi)}$ points ($\xi=\pm1$ being the hexagonal valley index), do not coincide any high-symmetry points of the rhombohedral BZ. The $2$D BZ associated with the projected $(111)$ plane of RG is shown above, which is hexagonal with the projections of the edge lines denoted by $\bar{K}^{(\xi)}$ at the corner points. The rhombohedral-to-hexagonal folding relation is illustrated in Fig. 2(b) by a vertical cut taken through suitable coincident high-symmetry points. Therefore, that $2$D BZ belonging to RG is identical to the $2$D BZ associated with the projected $(001)$ plane of stacks of $N$ layers. Such is a means extensively used, say, in angle-resolved photoemission spectroscopy,\cite{ohta07,ohta06} for displaying both the band structures and the the constant-energy contours of systems ranging from $2$D to $3$D.

A minimal tight-binding (TB) model, including only the nearest intralayer hopping $\beta_{0}$($=-2.73$ eV) and interlayer hopping $\beta_{1}$($=0.32$ eV), is described in Fig. 1(a) and used in this paper, where $\beta_{1}$ takes place between the vertically bonded bulk atoms in adjacent layers as shown in Fig. 1(a). The on-site Coulomb potential is not included here since it can be experimentally conditioned to be zero for realizing the QHEs, not to mention its diminishing and vanishing role for large $N$.

\par\noindent
\subsection{$3$D Dirac cone structure}
The $3$D Dirac cone structure in RG is described within the minimal TB model in the continuous approach as follows. Suppose that the low-energy bulk subbands in RG should be present in the vicinity of the $K^{(\xi)}$ edge lines,\cite{mcclure69} as in general stacks of graphene layers. Hence, we use the long-wavelength approximation for in-plane $(k_{x},k_{y})$ about the $K^{(\xi)}$ edge lines $(2\pi\xi/(\sqrt{3}a),2\pi\xi/(3a),k_{z})$, which are specified in Fig. 3(a). Based on the two TB Bloch functions $|A\rangle$ and $|B\rangle$ for the two bulk sublattices of RG, the Hamiltonian $\mathcal{H}^{(\xi)}$ with respect to $\xi$ is represented by a $2\times 2$ matrix, whose elements read
\begin{eqnarray}
H^{(\xi)}_{11}=H^{(\xi)}_{22}&=&0, \nonumber\\
H^{(\xi)}_{12}=H^{(\xi)*}_{21}&=&-\xi v_{0}p\exp{(-i\xi\varphi)}+%
\beta_{1}\exp{(ik_{z}d)},
\end{eqnarray}
where $p=\hbar(k_{x}^{2}+k_{y}^{2})^{1/2}$ is the in-plane momentum,
$\varphi=\arctan{(p_{y}/p_{x})}-7\pi/6$ is the azimuthal angle, and
$v_{0}=(3^{1/2}/2)a|\beta_{0}|/\hbar$ is the Fermi velocity as known below. The existence of DPs at the degeneracy points is expected in view of the chirality of $\mathcal{H}^{(\xi)}$.\cite{ho14} However, the location of the DP $(p_{D},\varphi_{D})$ disperses in the rhombohedral BZ and, in particular, varies continuously with $k_{z}$ due to the interlayer hopping $\beta_{1}$. That is, from Eq. (1), $(p_{D},\varphi_{D})$ is given by
\begin{equation}
p_{D}=\frac{\beta_{1}}{v_{0}}.
\end{equation}
\begin{equation}
\varphi_{D}=-\xi (k_{z}d-\frac{\pi}{2})-\frac{\pi}{2}.
\end{equation}
According to Eq.s (2) and (3), there are two distinguishable DP spirals with respect to $\xi=\pm1$. As $k_{z}$ increases from $-\pi/d$ to $\pi/d$ in the rhombohedral BZ, the azimuthal angle $\varphi_{D}$ decreases along the DP spiral with $\xi=1$ and increases with $\xi=-1$, making the clockwise and counterclockwise spiraling senses, respectively. It is straight to deduce that all the DPs have zero energy so that the Fermi surface shrinks to the DP spirals as a specialized result of the minimal model.\cite{heikkila11,ho13,mcclure69} As shown in Fig. 3(a), it is convenient to project the DP spirals onto the $2$D BZ for illustration, where the six valleys are distinguished by two indices $\xi=\pm1$ and two of them are specified. Several projected DPs around the specified $\bar{K}^{(+)}$ are marked in Fig. 3(b), for example. The DP spiral around the specified $K^{(-)}$ edge line can be similarly located using Eq.s (2) and (3). It is noted that those DPs outside the hexagon can be translated by reciprocal lattice vectors to other equivalent $K^{(\xi)}$ edge lines.

The Dirac cones are now expressed in terms of $(q, \vartheta)$ measured from the DPs for a fixed $k_{z}$. The coordinate is transformed by
$q^{2}=p^{2}+p_{D}^{2}-2p_{D}p\cos{(\varphi-\varphi_{D})}$ and
$\tan{[\vartheta+(1+\xi)\pi/2]}=(p\sin{\varphi}-p_{D}\sin{\varphi_{D}})%
(p\cos{\varphi}-p_{D}\cos{\varphi_{D}})^{-1}$. The Hamiltonian in Eq. (1) turns out to be transformed as
\begin{align}
\mathcal{H}^{(\xi)}=\xi v_{0}q\cos{\vartheta}\sigma_{x}+v_{0}q\sin{\vartheta}\sigma_{y},
\end{align}
where $\sigma_{x}$ and $\sigma_{y}$ are the Pauli matrices. The chiral Hamiltonian $\mathcal{H}^{(\xi)}$ described in Eq. (4) is just of the same form as monolayer graphene with the Fermi velocity $v_{0}$. It is remarked that the DP spirals should behave as nodal lines similar to the DPs as nodal points in the bulk of monolayer graphene.\cite{heikkila11,burkov11} The radius of the spiral location $\beta_{1}/v_{0}$ as described in Eq. (2) delimits the boundary of the bulk-surface correspondence, or the topological stability.\cite{heikkila11} Here we show that within the minimal model the $3$D Dirac cone structure in RG are composed of identical vertical and isotropic Dirac cones along the DP spirals. It suffices for the study of the evolution and dimensional crossover.

\par\noindent
\subsection{Chain model}
The lattice of ABCNG can be modelled as chains of atoms linked between the two surface sublattices, with the primitive unit set $\{B_{1},A_{1},B_{2},A_{2}\ldots,B_{N},A_{N}\}$. The chain model is applicable for arbitrary $N$, as a map from the continuous description.\cite{xiao11,guinea06} The mapped Hamiltonian $\mathcal{H}^{(N)}$ is based on $\{|B_{1}\rangle,|A_{N}\rangle,|A_{1}\rangle,|B_{2}\rangle,|A_{2}\rangle,|B_{3}\rangle,%
\ldots,|A_{N-1}\rangle,|B_{N}\rangle\}$, the set of the $2$D TB Bloch functions. The band structure is acquired in the $2$D BZ, which is associated with the projected $(001)$ plane of the stack. In the long-wavelength approximation about the high-symmetry corner points $\bar{K}^{(\xi)}$, referring to Fig. 3(a), $\mathcal{H}^{(N)}$ is represented as
\begin{equation}
\mathcal{H}^{(N)}=\left(
\begin{array}{cccccccccccc}
0 & 0 & v_{0}\pi^{\dag} & 0 & 0 &  0 & 0 & \ldots & 0 & 0 & 0 & 0\\
0 & 0 & 0 & 0 & 0 & 0 & 0 & \ldots & 0 & 0 & 0 & v_{0}\pi\\
v_{0}\pi & 0 & 0 & \beta_{1} & 0 & 0 & 0 & \ldots & 0 & 0 & 0 & 0\\
0 & 0 & \beta_{1} & 0 & v_{0}\pi^{\dag} & 0 & 0 & \ldots & 0 & 0 & 0 & 0\\
0 & 0 & 0 & v_{0}\pi & 0 & \beta_{1} & 0 & \ldots & 0 & 0 & 0 & 0\\
0 & 0 & 0 & 0 & \beta_{1} & 0 & v_{0}\pi^{\dag} & \ldots & 0 & 0 & 0 & 0\\
0 & 0 & 0 & 0 & 0 & v_{0}\pi & 0 & \ldots & 0 & 0 & 0 & 0\\
\vdots & \vdots & \vdots & \vdots & \vdots & \vdots & \vdots & \ddots & \vdots &\vdots & \vdots & \vdots\\
0 & 0 & 0 & 0 & 0 & 0 & 0 & \ldots & 0 & \beta_{1} & 0 & 0\\
0 & 0 & 0 & 0 & 0 & 0 & 0 & \ldots & \beta_{1} & 0 & v_{0}\pi^{\dag} & 0\\
0 & 0 & 0 & 0 & 0 & 0 & 0 & \ldots & 0 & v_{0}\pi & 0 & \beta_{1}\\
0 & v_{0}\pi^{\dag} & 0 & 0 & 0 & 0 & 0 & \ldots & 0 & 0 & \beta_{1} & 0
\end{array}%
\right)_{2N\times 2N},
\end{equation}
with in-plane momentum $\pi=-\xi p_{x}+ip_{y}$, $p_{x}\rightarrow p\cos{\varphi}$, $p_{y}\rightarrow p\sin{\varphi}$ being re-defined to fit the setting for the $3$D Dirac cone structure [Eq. (1)]. Without loss of generality for our purpose, we choose $\xi=1$ in the following. Since the bulk lattice of ABCNG has inversion symmetry, it can accommodate standing wave functions. This is a general property of systems having inversion or mirror symmetries, such as AB-stacked $N$-layer graphene with even or odd $N$, respectively. In the case of AB-stacking configuration, a direct zone-folding scheme along the $c$-axis can be applied, with definite $k_{z}$ wave numbers of the standing waves, so as to obtain the band structure for finite $N$ from AB-stacked graphite.\cite{koshino07,mak10} However, the situation of ABCNG is complicated due to the existing surface subbands.

The coupling between the surface and the bulk subbands is intuitively expected to decrease as $N$ increases. Within the infinite chain model, the surface layers are absent and thereby the $3$D Dirac cone structure should be derived as well. In Eq. (5), the bulk eigenequation of $\mathcal{H}^{(N)}$ given is expressed as
\begin{equation}
\left(
\begin{array}{cccc}
\beta_{1} & -\varepsilon & v_{0}\pi^{\dag} & 0\\
0 & v_{0}\pi & -\varepsilon & \beta_{1}
\end{array}%
\right)
\left(
\begin{array}{c}
\mathcal{U}_{A_{l-1}}\\
\mathcal{U}_{B_{l}}\\
\mathcal{U}_{A_{l}}\\
\mathcal{U}_{B_{l+1}}
\end{array}%
\right)
=0
\end{equation}
for arbitrarily layer label $l$, where $\varepsilon$ is the eigenenergy and $\mathcal{U}_{A_{l}}$ and $\mathcal{U}_{B_{l^{\prime}}}$ are the bulk components of the eigenmode. For infinite $N$, there are no boundaries and the labels $l$ are indistinguishable. Thus, the eigenmode takes the form
$(\mathcal{U}_{A_{l}},\mathcal{U}_{B_{l}})^{T}=%
(\tilde{\mathcal{U}}_{A_{l}},\tilde{\mathcal{U}}_{B_{l}})^{T}e^{i\kappa ld}$, where $\kappa$ belongs to the continuous set $\{\pm j\pi/(Nd)|j=1,2,\cdots\}_{N\to\infty}$. Inserting it into Eq. (6), the eigenequation in the subspace $(\tilde{\mathcal{U}}_{A_{l}},\tilde{\mathcal{U}}_{B_{l}})$ is obtained as
\begin{equation}
\left(
\begin{array}{cc}
0 & v_{0}\pi+\beta_{1}\exp{(i\kappa d)}\\
v_{0}\pi^{\dag}+\beta_{1}\exp{(-i\kappa d)} & 0
\end{array}%
\right)
\left(
\begin{array}{c}
\tilde{\mathcal{U}}_{A_{l}}\\
\tilde{\mathcal{U}}_{B_{l}}
\end{array}%
\right)
=\varepsilon
\left(
\begin{array}{c}
\tilde{\mathcal{U}}_{A_{l}}\\
\tilde{\mathcal{U}}_{B_{l}}
\end{array}%
\right).
\end{equation}
Evidently, Eq. (7) is equivalent to Eq. (1) for the bulk of RG if $\kappa$ is identified to be indexed by the $k_{z}$ wave number. In the bulk limit of ABCNG the bulk standing wave functions are thus characterized by definite wave vectors along the stacking direction (in the $c$-axis), as having been known.\cite{koshino10} Now that the surface layers are irrelevant, the direct zone-folding scheme is feasible. As shown in Fig 2(b), the $2$D BZ of the projected $(111)$ plane is just the cut of the folded hexagonal zone at $k_{z}=0$, where the whole $3$D Dirac cone structure are folded in.

\par\noindent
\subsection{Construction of the non-perturbative effective bulk Hamiltonian}
In order to understand the evolution and dimensional crossover, here we consider the finite chain model. We shall construct an effective Hamiltonian that is closed in the bulk subspace and embodied with the coupling to the surface subbands. The Hamiltonian $\mathcal{H}^{(N)}$ in Eq. (5) is partitioned as follows. The upper left $2\times2$ block is denoted by $H^{(N)}_{11}$ with respect to the surface subspace spanned by $(|B_{1}\rangle,|A_{N}\rangle)$, and the lower right $(2N-2)\times(2N-2)$ block is $H^{(N)}_{22}$ for the bulk subspace of the bulk sublattices. The coupling between $H^{(N)}_{11}$ and $H^{(N)}_{22}$ is present in the off-diagonal blocks $H^{(N)}_{12}$ and $H^{(N)}_{21}[=\bm{(}H^{(N)}_{12}\bm{)}^{\dag}]$. It is easy to identify the surface subbands as being lower in energy than the bulk subbands. The secular equation reduces to $\det{\bm{(}\mathcal{H}^{(N)}-\varepsilon\bm{)}}=\varepsilon^{2}(\varepsilon^{2}-\beta_{1}^{2})^{N-1}=0$ at $\bar{K}^{(+)}$, where the coupling is absent since $H^{(N)}_{12}$ and $H^{(N)}_{21}$ are zero matrices with $\pi=0$. The lowest eigenenergy $\varepsilon=0$ is associated with two degenerate eigenstates in the surface subspace, while the eigenenergies $\varepsilon=\pm\beta_{1}$ are each associated with $N-1$ degenerate eigenstates in the bulk subspace. Our goal is to construct an effective Hamiltonian closed in the bulk subspace. That is, a block diagonalization for the full Hamiltonian $\mathcal{H}^{(N)}$ is required:
\begin{equation}
\mathcal{H}^{(N)}=\left(
\begin{array}{cc}
H^{(N)}_{\mathrm{surf}} & 0\\
0 & H^{(N)}_{\mathrm{bulk}}
\end{array}%
\right),
\end{equation}
where $H^{(N)}_{\mathrm{surf}}$ and $H^{(N)}_{\mathrm{bulk}}$ are the effective Hamiltonians closed in the surface and the bulk subspaces, respectively. Considering Eq. (5) for $\mathcal{H}^{(N)}$ with $\pi\ne0$ in general, we expand its eigenvectors as $|\psi_{m^{\prime}}\rangle=\sum_{m=1}^{2N}C_{mm^{\prime}}|\psi_{m}^{(0)}\rangle$ in terms of the eigenvectors $|\psi_{m}^{(0)}\rangle$ of the uncoupled Hamiltonian ($\pi=0$). Hence, the Schr\"{o}dinger equation is reformulated to be
\begin{equation}
\left(
\begin{array}{cc}
H^{(N)}_{11}-\varepsilon & H^{(N)}_{12}\\
H^{(N)}_{21} & H^{(N)}_{22}-\varepsilon
\end{array}%
\right)\left(
\begin{array}{c}
C_{1}\\
C_{2}
\end{array}%
\right)=0,
\end{equation}
where $C_{1}$ and $C_{2}$ are, respectively, $2\times 2N$ and $(2N-2)\times 2N$ partitioned blocks of the matrix $\bm{[}C_{mm^{\prime}}\bm{]}$. The block diagonalization in Eq. (8) can be done by a similarity transformation of the basis from $\{|\psi_{m}^{(0)}\rangle\}$ to the unsolved set of $\{|\psi_{m^{\prime}}\rangle\}$. From Eq. (9) the effective surface Hamiltonian $H^{(N)}_{\mathrm{surf}}$ is given by
\begin{equation}
H^{(N)}_{\mathrm{surf}}(\varepsilon)=H^{(N)}_{11}%
-H^{(N)}_{12}\bm{(}H^{(N)}_{22}-\varepsilon\bm{)}^{-1}H^{(N)}_{21},
\end{equation}
and the effective bulk Hamiltonian $H^{(N)}_{\mathrm{bulk}}$ is
\begin{equation}
H^{(N)}_{\mathrm{bulk}}(\varepsilon)=H^{(N)}_{22}%
-H^{(N)}_{21}\bm{(}H^{(N)}_{11}-\varepsilon\bm{)}^{-1}H^{(N)}_{12},
\end{equation}
where the coupling effect is clearly expressed by both the second terms. Equivalently, Eq.s (10) and (11) can be derived using the Green's function $\mathcal{G}^{(N)}(\varepsilon)=\bm{(}\mathcal{H}^{(N)}-\varepsilon\bm{)}^{-1}$ such that one has $G^{(N)}_{11/22}(\varepsilon)=\bm{(}H^{(N)}_{\mathrm{surf/bulk}}-\varepsilon\bm{)}^{-1}$, where $G^{(N)}_{11}$ and $G^{(N)}_{22}$ are the partitioned blocks of the Green's function $\mathcal{G}^{(N)}$ can be obtained.\cite{allen86} An effective Hamiltonian can even be non-Hermitian in general, but this is not the case here. Moreover, it depend on energy $\varepsilon$. This can be deemed the price for the reduction in matrix dimension. The resulting eigenenergies and eigenmodes might also depend on energy $\varepsilon$.\cite{allen86} They ought to be in agreement with those of the true Hamiltonian at any level of energy $\varepsilon$ if the effective Hamiltonian is well constructed. Indeed, Eq.s (10) and (11) lead to the effective Hamiltonians for ABCNG, with energy dependences arising from the coupling between the surface and the bulk subbands.

The surface subbands have been well understood by the chiral effective Hamiltonian $H^{(3)}_{\mathrm{chiral}}$,\cite{koshino09,zhangf10} which was obtained in the framwork of Eq. (10) by retaining only the first order of the power series expansion in $\varepsilon/|H^{(N)}_{22}|$ for the coupling term before a renormalization. For arbitrary $N$, the chiral effective Hamiltonian was deduced to be\cite{min08}
\begin{equation}
H^{(N)}_{\mathrm{chiral}}=(\frac{-1}{\beta_{1}})^{N-1}\left(
\begin{array}{cc}
0 & (v_{0}\pi^{\dag})^{N}\\
(v_{0}\pi)^{N} & 0
\end{array}%
\right).
\end{equation}
It should be noted that Eq. (12) diverges outside the projection of the DP spiral with respect to $\bar{K}^{(+)}$, viz., $p>p_{D}=\beta_{1}/v_{0}$, referring to Eq. (2). The surface subbands inside are more and more flat and approach the zero energy with increasing $N$, as a consequence of bulk-surface correspondence.\cite{heikkila10}

For the bulk subbands, here we construct a non-perturbative effective Hamiltonian from Eq. (11) without using any perturbation procedure or power series expansion. In so doing we can reach the region around $p=p_{D}$, even though the bulk gap is, regarding the DP spiral, expected to be nearly but not exactly closed up in the bulk limit for arbitrarily large finite $N$. Therefore, the effective bulk Hamiltonian $H^{(N)}_{\mathrm{bulk}}(\varepsilon)$ is given by
\begin{equation}
H^{(N)}_{\mathrm{bulk}}(\varepsilon)=\left(
\begin{array}{cccccccccc}
\varepsilon^{-1}(v_{0}p)^{2} & \beta_{1} & 0 & 0 & 0 & \ldots & 0 & 0 & 0 & 0\\
\beta_{1} & 0 & v_{0}\pi^{\dag} & 0 & 0 & \ldots & 0 & 0 & 0 & 0\\
0 & v_{0}\pi & 0 & \beta_{1} & 0 & \ldots & 0 & 0 & 0 & 0\\
0 & 0 & \beta_{1} & 0 & v_{0}\pi^{\dag} & \ldots & 0 & 0 & 0 & 0\\
0 & 0 & 0 & v_{0}\pi & 0 & \ldots & 0 & 0 & 0 & 0\\
\vdots & \vdots & \vdots & \vdots & \vdots & \ddots & 0 & 0 & 0 & 0\\
0 & 0 & 0 & 0 & 0 & \ldots & 0 & \beta_{1} & 0 & 0\\
0 & 0 & 0 & 0 & 0 & \ldots & \beta_{1} & 0 & v_{0}\pi^{\dag} & 0\\
0 & 0 & 0 & 0 & 0 & \ldots & 0 & v_{0}\pi & 0 & \beta_{1}\\
0 & 0 & 0 & 0 & 0 & \ldots & 0 & 0 & \beta_{1} & \varepsilon^{-1}(v_{0}p)^{2}
\end{array}%
\right)_{(2N-2)\times (2N-2)}.
\end{equation}
Thus, the coupling effect on the bulk subbands manifests itself exactly at the two diagonal corners, with non-vanishing elements given by a parameter $(v_{0}p)^{2}/\varepsilon$ due to $H^{(N)}_{11}=0$. According to Eq. (13), the coupling is absent at $\bar{K}^{(+)}$ with $p=0$ and $H^{(N)}_{\mathrm{bulk}}(\varepsilon=\beta_{1})=H^{(N)}_{22}$ is obtained.

\par\noindent
\section{BULK SUBBANDS IN ABC-STACKED $N$-LAYER GRAPHENE}

\par\noindent
\subsection{Resolution for the secular equation}
For the bulk subbands in ABCNG, the eigenenergies of the effective bulk Hamiltonian $H^{(N)}_{\mathrm{bulk}}(\varepsilon)$ described in Eq. (13) are solved as follows. The secular equation $\det{\bm{(}H_{\mathrm{bulk}}^{(N)}(\varepsilon)-\varepsilon\bm{)}}=0$ is decomposed to be
\begin{equation}
f_{N}(\varepsilon)+\frac{(v_{0}p)^{2}}{\varepsilon}g_{N-1}(\varepsilon)=0,
\end{equation}
where $f_{N}(\varepsilon)[=\det{\bm{(}H_{22}^{(N)}-\varepsilon\bm{)}}]$ and $g_{N-1}(\varepsilon)$ are given by
\begin{align}
f_{N}(\varepsilon)&=\det{\left(
\begin{array}{cccccccccc}
-\varepsilon & \beta_{1} & 0 & 0 & 0 & \ldots & 0 & 0 & 0 & 0\\
\beta_{1} & -\varepsilon & v_{0}\pi^{\dag} & 0 & 0 & \ldots & 0 & 0 & 0 & 0\\
0 & v_{0}\pi & -\varepsilon & \beta_{1} & 0 & \ldots & 0 & 0 & 0 & 0\\
0 & 0 & \beta_{1} & -\varepsilon & v_{0}\pi^{\dag} & \ldots & 0 & 0 & 0 & 0\\
0 & 0 & 0 & v_{0}\pi & -\varepsilon & \ldots & 0 & 0 & 0 & 0\\
\vdots & \vdots & \vdots & \vdots & \vdots & \ddots & 0 & 0 & 0 & 0\\
0 & 0 & 0 & 0 & 0 & \ldots & -\varepsilon & \beta_{1} & 0 & 0\\
0 & 0 & 0 & 0 & 0 & \ldots & \beta_{1} & -\varepsilon & v_{0}\pi^{\dag} & 0\\
0 & 0 & 0 & 0 & 0 & \ldots & 0 & v_{0}\pi & -\varepsilon & \beta_{1}\\
0 & 0 & 0 & 0 & 0 & \ldots & 0 & 0 & \beta_{1} & -\varepsilon
\end{array}%
\right)}_{(2N-2)\times(2N-2)},
\end{align}
\begin{align}
g_{N-1}(\varepsilon)&=\det{\left(
\begin{array}{ccccccc}
-\varepsilon & v_{0}\pi^{\dag} & 0 & \ldots & 0 & 0 & 0\\
v_{0}\pi & -\varepsilon & \beta_{1} & \ldots & 0 & 0 & 0\\
0 & \beta_{1} & -\varepsilon & \ldots & 0 & 0 & 0\\
\vdots & \vdots & \vdots & \ddots & \vdots & \vdots & \vdots\\
0 & 0 & 0 & \ldots & -\varepsilon & v_{0}\pi^{\dag} & 0\\
0 & 0 & 0 & \ldots & v_{0}\pi & -\varepsilon & \beta_{1}\\
0 & 0 & 0 & \ldots & 0 & \beta_{1} & -\varepsilon
\end{array}
\right)}_{(2N-3)\times(2N-3)}.
\end{align}
Clearly, it is $g_{N-1}(\varepsilon)$ that arises from the coupling to the surface subbands. The relation between Eq.s (15) and (16), viz.,
\begin{align}
f_{N}(\varepsilon)&=-\beta_{1}^{2}f_{N-1}(\varepsilon)-\varepsilon g_{N-1}(\varepsilon), \notag\\
g_{N-1}(\varepsilon)&=-\varepsilon f_{N-1}(\varepsilon)-(v_{0}p)^{2} g_{N-2}(\varepsilon),\;\;\;\;\; N\ge3,
\end{align}
leads to a second-order recursive equation
\begin{equation}
f_{N}(\varepsilon)=(\lambda_{1}+\lambda_{2})f_{N-1}(\varepsilon)%
-\lambda_{1}\lambda_{2}f_{N-2}(\varepsilon),\;\;\;\;\; N\ge3,
\end{equation}
with
\begin{equation}
\lambda_{1}+\lambda_{2}=\varepsilon^{2}-\beta_{1}^{2}-(v_{0}p)^{2},\;\;\; \lambda_{1}\lambda_{2}=\beta_{1}^{2}(v_{0}p)^{2}.
\end{equation}
The initial conditions of $f_{N}(\varepsilon)$ is set forth in Eq. (15) as
\begin{equation}
f_{1}(\varepsilon)=1,\;\;\;f_{2}(\varepsilon)=\varepsilon^{2}-\beta_{1}^{2}-(v_{0}p)^{2}.
\end{equation}
The recursion of $g_{N-1}(\varepsilon)$ has the same form as $f_{N}(\varepsilon)$ in Eq. (18), with more tedious initial conditions.

In order to resolve $f_{N}(\varepsilon)$ with the initial conditions given in Eq. (20), we rewrite Eq. (18) as $f_{N}(\varepsilon)-\lambda_{1}f_{N-1}(\varepsilon)=\lambda_{2}\bm{(}f_{N-1}(\varepsilon)%
-\lambda_{1}f_{N-2}(\varepsilon)\bm{)}$. At first, we get $f_{N}(\varepsilon)-\lambda_{1}f_{N-1}(\varepsilon)=%
\lambda_{2}^{N-2}\bm{(}f_{2}(\varepsilon)-\lambda_{1}f_{1}(\varepsilon)\bm{)}$. Then we achieve the resolution and obtain $f_{N}(\varepsilon)=\lambda_{1}^{N-1}f_{1}(\varepsilon)%
+\sum^{N}_{\nu=2}\lambda_{1}^{N-\nu}\lambda_{2}^{\nu-2}\bm{(}f_{2}(\varepsilon)-\lambda_{1}f_{1}(\varepsilon)\bm{)}$ for arbitrary $N$, which is written, for conciseness, as
\begin{equation}
f_{N}(\varepsilon)=\frac{1}{\lambda_{1}-\lambda_{2}}[(\lambda_{1}^{N-1}-\lambda_{2}^{N-1})f_{2}(\varepsilon)%
-\lambda_{1}\lambda_{2}(\lambda_{1}^{N-2}-\lambda_{2}^{N-2})f_{1}(\varepsilon)],\;\;\;\;\; N\ge3.
\end{equation}
The resolved expression of $g_{N-1}(\varepsilon)$ can be obtained with Eq. (21) according to Eq.s (17).

\par\noindent
\subsection{Eigenenergy spectrum}
In proceeding to solve the eigenvalues of $H_{\mathrm{bulk}}^{(N)}(\varepsilon)$ from the secular equation Eq. (14), $f_{N}(\varepsilon)$ and $g_{N-1}(\varepsilon)$ obtained from Eq.s (17) and (21) are manipulated further. The variables $\lambda_{1}$ and $\lambda_{2}$ defined in Eq. (19) form a complex conjugate pair if and only if
\begin{equation}
(\lambda_{1}+\lambda_{2})^{2}-4\lambda_{1}\lambda_{2}=[\varepsilon-(\beta_{1}+v_{0}p)][\varepsilon-(\beta_{1}-v_{0}p)]%
[\varepsilon+(\beta_{1}+v_{0}p)][\varepsilon+(\beta_{1}-v_{0}p)]<0.
\end{equation}
That is, the whole eigenenergy spectrum $\varepsilon(p)$ are enveloped by four cone-like branches: $\varepsilon-(\beta_{1}\pm v_{0}p)=0$ and $\varepsilon+(\beta_{1}\pm v_{0}p)=0$. The intersections of these branches at $(p=p_{D}=\beta_{1}/v_{0},\varepsilon=0)$ are just the projection of the DP spiral on the $2$D BZ [Eq. (2)], referring to Fig. 3(b). This premise is definitely valid in view of the known band structure of ABCNG.\cite{koshino10,nakamura08} By changing variables as $\lambda_{1}=\rho e^{i\phi}$ and $\lambda_{2}=\rho e^{-i\phi}$ in Eq. (19), viz.,
\begin{equation}
\rho=\beta_{1}v_{0}p,\;\;\; \cos{\phi}=\frac{\varepsilon^{2}-\beta_{1}^{2}-(v_{0}p)^{2}}{2\beta_{1}v_{0}p},
\end{equation}
Eq. (21) is transformed to be
\begin{equation}
f_{N}(\varepsilon)=\frac{\rho^{N-1}}{\sin{\phi}}\sin{N\phi},\;\;\;\;\;N\ge3,
\end{equation}
using the trigonometric identity $2\cos{\phi}\sin{(N-1)\phi}-\sin{(N-2)\phi}=\sin{N\phi}$. Moreover, Eq. (24) together with Eq. (17) yields
\begin{equation}
g_{N-1}(\varepsilon)=-\frac{\beta_{1}\rho^{N-2}}{\varepsilon\sin{\phi}}%
\bm{(}v_{0}p\sin{N\phi}+\beta_{1}\sin{(N-1)\phi}\bm{)},\;\;\;\;\;N\ge3.
\end{equation} In terms of Eq.s (24) and (25) and in use of Eq. (23), the secular Eq. (14) becomes
\begin{equation}
\frac{\beta_{1}^{2}\rho^{N-1}}{\varepsilon^{2}\sin{\phi}}%
\bm{(}\sin{N\phi}+\frac{v_{0}p}{\beta_{1}}\sin{(N+1)\phi}\bm{)}=0,\;\;\;\;\; N\ge3,
\end{equation}
where energy $\varepsilon$ turns out to be factored out in spite of the $\varepsilon$-dependent Hamiltonian $H_{\mathrm{bulk}}^{(N)}(\varepsilon)$. It is noted that the coupling to the surface subbands has a manifestation in the second term in Eq. (26). There should be $N-1$ roots to Eq. (26) with respect to the $2N-2$ eigenvalues of $H_{\mathrm{bulk}}^{(N)}(\varepsilon)$. Once those roots $\phi_{j}(p)$, $j=1,2,\ldots,N-1$, are determined as functions of $p$, the eigenenergies of $H_{\mathrm{bulk}}^{(N)}(\varepsilon)$ can be acquired according to Eq. (23). That is,
\begin{equation}
\varepsilon^{(\pm)}_{j}(p)=\pm[\beta_{1}^{2}+(v_{0}p)^{2}+2\beta_{1}v_{0}p\cos{\phi_{j}(p)}]^{1/2},\;\;\;j=1,2,\ldots,N-1,
\end{equation}
where $\pm$ refer to the conduction and the valence bulk subbands, respectively.

The roots $\phi_{j}(p)$ to Eq. (26) index the bulk subbands. To find out $\phi_{j}(p)$ from Eq. (26), we firstly survey $v_{0}p/\beta_{1}=0$ at $\bar{K}^{(+)}$ at $p=0$ and $v_{0}p/\beta_{1}=1$ at $p=p_{D}$. For $v_{0}p/\beta_{1}=0$, the roots $\phi_{j}$ to Eq. (26) are simply specified by $\sin{N\phi}=0$, $N\ge3$, given by
\begin{equation}
N\phi_{j}=j\pi,\;\;\;j=1,2,\ldots,N-1.
\end{equation}
For $v_{0}p/\beta_{1}=1$, Eq. (26) becomes
$[\beta_{1}^{2}\rho^{N-1}/\bm{(}\varepsilon^{2}\sin{(\phi/2)}\bm{)}]\sin{(N+1/2)\phi}=0$, $N\ge3$, and the roots $\phi_{j}$ are solved as
\begin{equation}
N\phi_{j}=j\pi-\frac{1}{2}\phi_{j},\;\;\;j=1,2,\ldots,N-1.
\end{equation}
In Eq.s (28) and (29), the numerals $j$ are so specified as to exclude the case of $\sin{\phi}=0$. The two sets of $\phi_{j}$ at the two ends of $v_{0}p/\beta_{1}$ are different. Between these two ends, the bulk indices $\phi_{j}(p)$ are implicit in Eq. (26) and can be solved iteratively by
\begin{equation}
N\phi_{j}=j\pi-\theta_{j}(p,\phi_{j}),\;\;\;j=1,2,\ldots,N-1,
\end{equation}
as roots to $\sin{(N\phi+\theta)}=0$, where $\tan{\theta}=\sin{\phi}[\beta_{1}/(v_{0}p)+\cos{\phi}]^{-1}$ is defined. Here we note that Eq. (26) agrees with previous researches,\cite{koshino10} whose derivation is conducted by simultaneously considering the surface and the bulk subbands with boundary conditions for the bulk wave functions fixed \textit{a priori} at fictitious atoms outside of ABCNG. In distinction, the present derivation of Eq. (26) is based on the effective bulk Hamiltonian and, therefore, the coupling effect has a definite manifestation. According to Eq. (26), the coupling effect is measured by $v_{0}p/\beta_{1}$, the distance from $\bar{K}^{(+)}$ with $p=0$ to the projection of the DP spiral with $p_{D}$ [Eq. (2)]. As revealed in Eq. (30), $\theta_{j}(p,\phi_{j})$ varies monotonically from $0$ to $\phi_{j}/2$. For large $N$, $N\phi_{j}(\approx j\pi)$ dominates in Eq. (30), indicating that the coupling effect is weakened.

\par\noindent
\subsection{Characterization of the eigenmodes}
In association with the eigenenergies $\varepsilon^{(\pm)}_{j}(p)$ of the effective bulk Hamiltonian $H_{\mathrm{bulk}}^{(N)}(\varepsilon)$, now the eigenmodes are characterized with respect to $\phi_{j}$. According to the structure of $H_{\mathrm{bulk}}^{(N)}(\varepsilon)$, the inherent recursive relation among the $\beta_{1}$-bonded components of an eigenmode is given by
\begin{equation}
\left(
\begin{array}{cc}
0 & -v_{0}\pi^{\dag}\\
\beta_{1} & -\varepsilon
\end{array}%
\right)
\left(
\begin{array}{c}
\mathcal{U}_{B_{l+1}}\\
\mathcal{U}_{A_{l}}
\end{array}%
\right)
=\left(
\begin{array}{cc}
-\varepsilon & \beta_{1}\\
-v_{0}\pi & 0
\end{array}%
\right)
\left(
\begin{array}{c}
\mathcal{U}_{B_{l}}\\
\mathcal{U}_{A_{l-1}}
\end{array}%
\right),\;\;\;l=2,3,\ldots,N-1,
\end{equation}
for which $\mathcal{\underline{U}}_{l}=(\mathcal{U}_{B_{l+1}},%
\mathcal{U}_{A_{l}})^{T}$ is defined below. The recursion Eq. (31) is equivalent to
\begin{equation}
\mathcal{\underline{U}}_{l}=R\mathcal{\underline{U}}_{l-1},%
\;\;\;\mathcal{\underline{U}}_{l}=R^{l-1}\mathcal{\underline{U}}_{1}%
\;\;\;l=2,3,\ldots,N-1,
\end{equation}
with $R$ obtained as
\begin{equation}
R=\frac{1}{\beta_{1}v_{0}\pi^{\dag}}\left(
\begin{array}{cc}
\varepsilon^{2}-(v_{0}p)^{2} & -\varepsilon\beta_{1}\\
\varepsilon\beta_{1} & -\beta_{1}^{2}
\end{array}%
\right).
\end{equation}
A deduction of Eq. (32) leads to
\begin{equation}
\mathcal{\underline{U}}_{N-1}=R^{N-2}\mathcal{\underline{U}}_{1}.
\end{equation}
Here we put the boundary conditions due to the coupling to the surface subbands as known from $H_{\mathrm{bulk}}^{(N)}(\varepsilon)$ in Eq. (13), that is,
\begin{equation}
\mathcal{\underline{U}}_{1}=\mathcal{U}_{A_{1}}\left(
\begin{array}{c}
\beta_{1}^{-1}(\varepsilon-\sigma)\\
1
\end{array}%
\right),\;\;\;%
\mathcal{\underline{U}}_{N-1}=\mathcal{U}_{B_{N}}\left(
\begin{array}{c}
1\\
\beta_{1}^{-1}(\varepsilon-\sigma)
\end{array}%
\right),
\end{equation}
where $\sigma=(v_{0}p)^{2}/\varepsilon$ refers to the coupling elements of $H_{\mathrm{bulk}}^{(N)}(\varepsilon)$.

For the power $R^{N-2}$ in Eq. (34), we shall derived an explicit expression in terms of $R$'s eigenvalues, $\tilde{\lambda}_{1}$ and $\tilde{\lambda}_{2}$, by the aid of a general theorem for non-degenerate eigenvalues,\cite{perlis56} instead of performing a direct but tedious calculation with the similarity transformation. Let $h(R)$ denote any polynomial of $R$. This theorem applied here sets forth
\begin{equation}
h(R)=\sum_{i=1,2}P_{i}(R)h(\tilde{\lambda}_{i}),
\end{equation}
where $P_{i}(R)$ is given by
\begin{equation}
P_{i}(R)=\frac{R-\tilde{\lambda}_{j}I}{\tilde{\lambda}_{i}-\tilde{\lambda}_{j}},%
\;\;\;j\neq i,
\end{equation}
with $I$ being the identity matrix. It is noted that $P_{i}(R)$ is the matrix that projects any state to the $i$th eigenstate of $R$. Owing to Eq.s (36) and (37) any power $R^{l}$ can be expressed as
\begin{equation}
R^{l}=(\tilde{\lambda}_{1}-\tilde{\lambda}_{2})^{-1}%
[(\tilde{\lambda}_{1}\tilde{\lambda}_{2}^{l}-\tilde{\lambda}_{1}^{l}\tilde{\lambda}_{2})I%
+(\tilde{\lambda}_{1}^{l}-\tilde{\lambda}_{2}^{l})R].
\end{equation}

From Eq. (33) the eigenvalues of $R$ turn out to be related to $\lambda_{1}$ and $\lambda_{2}$ given in Eq. (19), viz., $\tilde{\lambda}_{i}=\lambda_{i}/(\beta_{1}v_{0}\pi^{\dag})$,$i=1,2$. Referring to Eq. (5) for $\pi^{\dag}=pe^{i\varphi^{\prime}}$, with $\varphi^{\prime}=\varphi+\pi$ in consistency with Eq. (1), and changing variables for $\lambda_{1,2}$ as in Eq. (23), $\tilde{\lambda}_{1,2}$ are obtained as
\begin{equation}
\tilde{\lambda}_{1}=\exp{\bm{(}i(\phi-\varphi^{\prime})\bm{)}},%
\;\;\;\tilde{\lambda}_{2}=\exp{\bm{(}-i(\phi+\varphi^{\prime})\bm{)}},
\end{equation}
and
\begin{equation}
\tilde{\lambda}_{1}+\tilde{\lambda}_{1}=2\exp{(-i\varphi^{\prime})}\cos{\phi},%
\;\;\;\tilde{\lambda}_{1}\tilde{\lambda}_{2}=\exp{(-2i\varphi^{\prime})}.
\end{equation}
For convenience below, we define and use $\eta_{1,2}=\lambda_{1,2}/\rho=e^{\pm i\phi}$, respectively. As a result, Eq. (38) becomes
\begin{equation}
R^{l}=\frac{\exp{(-il\varphi^{\prime})}}{\eta_{1}-\eta_{2}}\left[\left(
\begin{array}{cc}
\eta_{1}^{l+1}-\eta_{2}^{l+1} & 0\\
0 & \eta_{2}^{l-1}-\eta_{1}^{l-1}
\end{array}%
\right)+
\frac{\eta_{1}^{l}-\eta_{2}^{l}}{v_{0}p}\left(
\begin{array}{cc}
\beta_{1} & -\varepsilon\\
\varepsilon & -\beta_{1}
\end{array}%
\right)
\right],
\end{equation}
where we have change the element $R_{11}=\varepsilon^{2}-(v_{0}p)^{2}$ [Eq. (33)] to be $\lambda_{1}+\lambda_{2}+\beta_{1}^{2}$ with Eq. (23).

All the eigenmodes must be characterized, before being solved out, by requiring the non-trivial solutions as one put Eq. (41) into Eq. (34) with the boundary conditions Eq. (35). The characterization is equivalent to the determination of the eigenenergies $\varepsilon^{(\pm)}_{j}(p)$ with Eq. (26), which is derived from the secular equation Eq. (14). The non-trivial condition is thus given by
\begin{equation}
(\frac{\sigma-\varepsilon}{\beta_{1}})^{2}(\eta_{1}^{N-1}-\eta_{2}^{N-1})+%
\frac{\sigma^{2}-\varepsilon^{2}+\beta_{1}^{2}}{\beta_{1}v_{0}p}%
(\eta_{1}^{N-2}-\eta_{2}^{N-2})+\eta_{1}^{N-3}-\eta_{2}^{N-3}=0.
\end{equation}
By using Eq. (23) and the identity $(\eta_{1}+\eta_{2})(\eta_{1}^{l}-\eta_{2}^{l})=%
\eta_{1}^{l+1}-\eta_{2}^{l+1}+\eta_{1}^{l-1}-\eta_{2}^{l-1}$ repeatedly, it is straight to reduce Eq. (42) to
\begin{equation}
\frac{\beta_{1}v_{0}p}{\varepsilon^{2}}[\eta_{1}^{N}-\eta_{2}^{N}+%
\frac{v_{0}p}{\beta_{1}}(\eta_{1}^{N+1}-\eta_{2}^{N+1})]=0.
\end{equation}
Obviously, Eq. (43) is equivalent to Eq. (26) owing to the identity $\eta_{1}^{\nu}-\eta_{2}^{\nu}=2i\sin{\nu\phi}$.

In terms of the characterized $\phi_{j}$, the eigenmode $\mathcal{\underline{U}}_{j}$ can be obtained from  Eq.s (32), (35) and (41), with components $\mathcal{\underline{U}}_{j,l}$, $l=1,2,\ldots,N-1$, given by
\begin{equation}
\left(
\begin{array}{c}
\mathcal{U}_{j,B_{l+1}}\\
\mathcal{U}_{j,A_{l}}
\end{array}%
\right)
=\frac{\mathcal{U}_{j,A_{1}}\exp{\bm{(}-i(l-1)\varphi^{\prime}\bm{)}}}{\sin{\phi_{j}}}\left(
\begin{array}{c}
\varepsilon^{-1}\beta_{1}\bm{(}\sin{l\phi_{j}}+%
\beta_{1}^{-1}v_{0}p\sin{(l+1)\phi_{j}}\bm{)}\\
\sin{l\phi_{j}}
\end{array}%
\right).
\end{equation}
The energy ($\varepsilon$) dependence of the relative magnitudes of $\mathcal{\underline{U}}_{j,l}$, shown as $\beta_{1}/\varepsilon$ in Eq. (44), is a reasonable result of the $\varepsilon$-dependent effective bulk Hamiltonian $H_{\mathrm{bulk}}^{(N)}(\varepsilon)$. At any given level of energy $\varepsilon$, the eigenmode $\mathcal{\underline{U}}_{j}$ can be evaluated from Eq. (44). For verification, $\mathcal{\underline{U}}_{j,N-1}$ is given by
\begin{equation}
\left(
\begin{array}{c}
\mathcal{U}_{j,B_{N}}\\
\mathcal{U}_{j,A_{N-1}}
\end{array}%
\right)
=\frac{\mathcal{U}_{j,A_{1}}\exp{\bm{(}-i(N-2)\varphi^{\prime}\bm{)}}}{\sin{\phi_{j}}}\left(
\begin{array}{c}
\varepsilon(v_{0}p)^{-1}\sin{N\phi_{j}}\\
\bm{(}\varepsilon^{2}-(v_{0}p)^{2}\bm{)}(\beta_{1}v_{0}p)^{-1}\sin{N\phi_{j}}
\end{array}%
\right),
\end{equation}
which is simplified owing to $\sin{N\phi_{j}}+(v_{0}p/\beta_{1})\sin{(N+1)\phi_{j}}=0$ [Eq. (26)] and can be shown to satisfy the boundary condition Eq. (35). It is easy to extend Eq. (44) to $l=0$ and $l=N$, so that the present characterization of $\phi_{j}$ with respect to Eq. (26) proves to agree with the \textit{a priori} imposition of $\mathcal{U}_{j,A_{0}}=0$ and $\mathcal{U}_{j,B_{N+1}}=0$ at fictitious bulk atoms outside of ANCNG.\cite{koshino10} The eigenmodes $\mathcal{\underline{U}}_{j}$ constitute standing wave functions, of which the bulk indices $\phi_{j}$ are related to the $k_{z}$ wave numbers by $k_{z}=\pm\phi_{j}/d$. Because of the coupling to the surface subbands, $\phi_{j}$ is a function of $p$, implicitly given in Eq. (30). Therefore, in ABCNG the wave length of the standing wave across the stack of graphene layers depends on the in-plane Bloch momentum.

\par\noindent
\subsection{Bulk density of states}
The bulk DOS is a sum over all the local DOS of individual bulk sublattices. Generally, the local DOS is related to the Green's function in the bulk subspace. If $\varepsilon\rightarrow \varepsilon+i0^{+}$ is made, the Green's function associated with the effective bulk Hamiltonian reads $G^{(N)}_{22}(\varepsilon+i0^{+})=\bm{(}H^{(N)}_{\mathrm{bulk}}(\varepsilon)-\varepsilon-i0^{+}\bm{)}^{-1}$.\cite{allen86} Referring to Eq. (27), the subband index $j(=1,2,\ldots,N-1)$ and band index $s(\stackrel{\mathrm{def}}{=}\pm)$ are lumped as $m(\in \{j\}\otimes\{ s\})$ in the following. If the bulk subspace is spanned by the set of eigenvectors $\{|\psi_{m}\rangle\}$ of $H^{(N)}_{\mathrm{bulk}}(\varepsilon)$, with the eigenvalues $\{\varepsilon_{m}\}$, the Green's function represented as $\sum_{m}|\psi_{m}\rangle\langle\psi_{m}|%
\bm{(}H^{(N)}_{\mathrm{bulk}}(\varepsilon)-\varepsilon-i0^{+}\bm{)}^{-1}|\psi_{m}\rangle\langle\psi_{m}|$ reduces to
\begin{equation}
G^{(N)}_{22}(\varepsilon+i0^{+})=-\sum_{m}%
\frac{|\psi_{m}\rangle\langle\psi_{m}|}{\varepsilon-\varepsilon_{m}+i0^{+}},
\end{equation}
for which the identity $\langle\psi_{m}|\mathcal{O}|\psi_{m}\rangle^{-1}=\langle\psi_{m}|\mathcal{O}^{-1}|\psi_{m}\rangle$ for an operator $\mathcal{O}$ is used. Based on the $2N-2$ TB Bloch functions, the imaginary part of the diagonal element $g_{nn}$ of $G^{(N)}_{22}(\varepsilon+i0^{+})$ in Eq. (46) is given by
\begin{equation}
\mathrm{Im}\; g_{nn}=\pi\sum_{m}|c_{mn}|^{2}\delta(\varepsilon-\varepsilon_{m}),\;\;\;\;\; n=1,2,\ldots,2N-2,
\end{equation}
with $c_{mn}=\langle\phi_{n}|\psi_{m}\rangle$ being the component of $|\psi_{m}\rangle$ at the $n$th bulk sublattice, where $\pi\delta(t)=w/(t^{2}+w^{2})$, $w\rightarrow 0$, is defined for the delta function. The local DOS has a manifestation in Eq. (47) since the eigenvector $|\psi_{m}\rangle$ contributes a probability density $|c_{mn}|^{2}$ at the $n$th bulk sublattice. In the infinitely extended $(x,y)$ plane, the number of states is obtained by counting the allowed wave vectors in $(k_{x},k_{y})$ space. Hence, the local DOS is given by
\begin{equation}
D^{(N)}_{nn}(\varepsilon)=\frac{1}{\pi}\int_{\mathrm{BZ}} \frac{d\mathbf{k}}{(2\pi)^{2}}\mathrm{Im}\; g_{nn},
\end{equation}
where the integration turns out to run along the circular isoenergetic path with respect to each subband $\varepsilon_{m}(p=\hbar k)$ given in Eq. (27).

The bulk DOS is obtained from the local DOS in Eq. (48) by summing $D^{(N)}_{nn}(\varepsilon)$ over all the $2N-2$ bulk sublattices, given by $D^{(N)}_{\mathrm{bulk}}(\varepsilon)=\sum^{2N-2}_{n}D^{(N)}_{nn}(\varepsilon)$. This leads to
\begin{equation}
D^{(N)}_{\mathrm{bulk}}(\varepsilon)=\sum_{m}\int_{\mathrm{BZ}} \frac{d\mathbf{k}}{(2\pi)^{2}}\delta\bm{(}\varepsilon-\varepsilon_{m}(k)\bm{)},
\end{equation}
where $\sum^{2N-2}_{n}|c_{mn}|^{2}=1$ for normalized $|\psi_{m}\rangle$. In the calculation with Eq. (49), we use the Lorentzian $(\Gamma/\pi)[\bm{(}\varepsilon-\varepsilon_{m}(k)\bm{)}^{2}+\Gamma^{2}]^{-1}$ with a small width $\Gamma$ to approximate the delta function $\delta\bm{(}\varepsilon-\varepsilon_{m}(k)\bm{)}$.

\par\noindent
\section{RESULTS AND ANALYSES}

\par\noindent
\subsection{Evolution}
The eigenenergies $\varepsilon^{(\pm)}_{j}(p)$ of the effective bulk Hamiltonian $H^{(N)}_{\mathrm{bulk}}(\varepsilon)$ of ABCNG are calculated from Eq.s (27) and (30) for various numbers ($N$) of layers. The results are presented and discussed below, accompanied with the energies of the surface subbands
\begin{equation}
\frac{\varepsilon}{\beta_{1}}=\pm(\frac{v_{0}p}{\beta_{1}})^{N},
\end{equation}
which is obtained from the chiral effective Hamiltonian $H^{(N)}_{\mathrm{chiral}}$ given in Eq. (12). In Fig. 4, the evolution of the band structure of ABCNG is clear in an overview from a few layers ($N=3,4,\ldots,8$) to a lot of layers ($N\approx100$). Certainly, the surface subbands are rapidly flatten inside the interval, $0<v_{0}p/\beta_{1}<1=v_{0}p_{D}/\beta_{1}$, between $\bar{K}^{(+)}$ and the projection of the DP spiral.\cite{heikkila10} As having been noted, outside the interval the energies of $H^{(N)}_{\mathrm{chiral}}$ diverge dramatically. The actual surface subbands are, however, known to be suppressed due to band repulsion, and can be acquired using the full Hamiltonian consisting of coupled surface and bulk states over the whole BZ.\cite{nakamura08,lin14} By contrast, our results from the non-perturbative $H_{\mathrm{bulk}}^{(N)}(\varepsilon)$ agree with those full Hamiltonian results over a wide region.

The intricate bulk subbands $\varepsilon^{(\pm)}_{j}(p)$ are analytically unraveled at first. The eigenenergy spectrum has isotropic and electron-hole symmetries, as a consequence of the minimal model. All the $N-1$ conduction bulk subbands $\varepsilon^{(+)}_{j}(p)$, as well as all the $N-1$ valence bulk subbands $\varepsilon^{(-)}_{j}(p)$, are degenerate at $\bar{K}^{(+)}$ ($p=0$), where $\phi_{j}$ given in Eq. (28 can be paired according to $\cos{(j\pi/N)}=-\cos{\bm{(}(N-j)\pi/N\bm{)}}$, with $\phi_{N/2}$ standing solely for $\cos{\phi_{N/2}}=0$ if $N$ is even. Near $\bar{K}^{(+)}$ the conduction and the valence bulk subbands comprise pairs of $\varepsilon^{(+)}_{i\pm}(p)\approx\beta_{1}\pm v_{0}p\cos{(i\pi/N)}$ and pairs of $\varepsilon^{(-)}_{i\pm}(p)\approx-\beta_{1}\pm v_{0}p\cos{(i\pi/N)}$, respectively, with $i=1,2,\ldots,\lfloor(N-1)/2\rfloor$, where $\lfloor \cdot \rfloor$ denotes the integer part of $\cdot$. Besides, for even $N$ they also comprise quadratic subbands $\varepsilon^{(+)}_{i=N/2}(p)\approx\beta_{1}+(v_{0}p)^{2}/2\beta_{1}$ and $\varepsilon^{(-)}_{i=N/2}(p)\approx-\beta_{1}-(v_{0}p)^{2}/2\beta_{1}$, respectively. In each of the $\lfloor(N-1)/2\rfloor$ pairs of conduction (valence) bulk subbands, it is just $\varepsilon^{(+)}_{i-}(p)=\varepsilon^{(+)}_{j=N-i}(p)$ [$\varepsilon^{(-)}_{i+}(p)=\varepsilon^{(-)}_{j=N-i}(p)$] that extends downward (upward) from $\bm{(}p=0,\varepsilon=\beta_{1}(-\beta_{1})\bm{)}$. As $p$ goes from $p=0$ toward $p=p_{D}$, $\phi_{j}$ deviates from $j\pi/N$ due to the coupling to the surface subbands, and is calculated by using Eq. (30). This subband turns upward (downward) and form concave-up (concave-down) valley around $p_{D}$ as shown in Fig. 4. Such valleys are annular with egdes around $p_{D}$ since all the eigenenergies are isotropic. The turning is due to the nonlinear terms of $v_{0}p/\beta_{1}$ in the power expansion of $\varepsilon^{(\pm)}_{j}(p)$ in Eq. (27).

The evolution of the bulk subbands is investigated as follows. There is a bulk energy gap between the valley edges of the lowest concave-up conduction and the highest concave-down valence bulk subbands $\varepsilon^{(\pm)}_{j=N-1}(p)$. By differentiating $\varepsilon^{(\pm)}_{j=N-1}(p)$ in Eq. (27) with respect to $v_{0}p$ near $v_{0}p_{D}$, the edge momentum in ABCNG is approximately given by \begin{equation}
p_{\mathrm{edge}}^{(N)}=\frac{\beta_{1}}{v_{0}}\cos{(\frac{\pi}{2N-1})}.
\end{equation}
The edge energies $\varepsilon^{(\pm)}_{j=N-1}(p_{\mathrm{edge}}^{(N)})$ are then obtained, and so is the bulk gap
\begin{equation}
\Delta_{\mathrm{gap}}^{(N)}=4\beta_{1}\sin{(\frac{\pi}{2N-1})}.
\end{equation}
In the course of evolution, Eq.s (51) and (52) dictate that $p_{\mathrm{edge}}^{(N)}$ increases and approaches $p_{D}=\beta_{1}/v_{0}$ while $\Delta_{\mathrm{gap}}^{(N)}$ decreases and approaches $0$. The results displayed in Fig. 4 show that $p_{\mathrm{edge}}^{(N)}$ evolves rapidly, but comparatively $\Delta_{\mathrm{gap}}^{(N)}$ evolves slowly. By contrast, in AB-stacked $N$-layer graphene the band structure exhibits similar $3$D features to AB-stacked graphite with just several layers ($N\approx10$).\cite{partoens06} Taking the advantage of the expression of Eq. (27), the bulk subbands $\varepsilon^{(\pm)}_{j}(p)$ can be calculated for arbitrary $N$. As observed in Fig. 4, the bulk gap $\Delta_{\mathrm{gap}}^{(N)}$ at $N\approx100$ has become narrow but still open.

The bulk DOSs $D^{(N)}_{\mathrm{bulk}}(\varepsilon)$ in ABCNG ranging from a few to one hundred layers are also calculated and shown in Fig. 5. There are as many peaks in $D^{(N)}_{\mathrm{bulk}}(\varepsilon)$ as there are the bulk subbands. Specifically, each bulk subband yields a peak at its own valley edge. The most prominent two peaks arise from $\varepsilon^{(\pm)}_{j=N-1}(p)$. They are separated by the aforementioned decreasing bulk gap $\Delta_{\mathrm{gap}}^{(N)}$ and hence slowly approach each other in the evolution. Besides, a dip is present at energy $\beta_{1}$ ($-\beta_{1}$) in the conduction (valence) bulk subband spectrum. This reflects the cusp at $\bar{K}^{(+)}$, where all the subbands are degenerate. All these features are gradually smeared as $N$ increases. by taking the infinite limit of $N$

\par\noindent
\subsection{Dimensional crossover}
The dimensional crossover from the bulk subbands in ABCNG to the $3$D Dirac cone structure in RG is elucidated now. Recall that in Eq. (30) the bulk index $\phi_{j}$ is implicitly given by $N\phi_{j}=j\pi-\theta_{j}(p,\phi_{j})$ with $j=1,2,\ldots,N-1$, where the functional $\theta_{j}(p,\phi_{j})$ arising from the coupling to the surface subbands varies from $0$ at $p=0$ [Eq. (28)] to $\phi_{j}/2$ at $p=p_{D}$ [Eq. (29)]. For finite $N$, the presence of the bulk gap is caused by $\phi_{N-1}(p_{D})=(N-1)\pi/(N+1/2)$ for $\cos{\phi_{j}(p)}$ in Eq. (27). In the bulk limit, $\theta_{j}(p,\phi_{j})$ is negligible compared with $N\phi_{j}$. Hence, we acquire a continuous set $\{\phi_{j}=j\pi/N|j=1,2,\ldots,N-1\}$ if $N$ approaches infinity. The relation of $\phi_{j}(p)$ to the $k_{z}$ wave number of the standing wave functions in ABCNG [Eq.(44)] has been given by $k_{z}=\pm\phi_{j}(p)/d$ by characterizing the eigenmodes. Therefore, $k_{z}\in[-\pi/d,\pi/d]$ is founded in connection to the continuous set of $\phi_{j}$ in the infinite limit of $N$. Now that $k_{z}$ is definite, being irrelevant to the in-plane momentum $p$ in the absence of the coupling effect, it is feasible to do the direct zone folding along the $c$-axis as shown in Fig. 2(b). The whole $3$D Dirac cone structure is thus folded in the $k_{z}=\lim_{N\to\infty}[\pm j\pi/(Nd)]=0$ cut of the hexagonal zone that is folded from the rhombohedral BZ. As shown in Fig. 2(b), the $k_{z}=0$ cut is equivalent to the $2$D BZ.

In the infinite limit of $N$, the set $\{i|i=1,2,\cdots,\lfloor(N-1)/2\rfloor\}$ contains an infinite subset such that its elements $i^{\prime}$ render $\phi_{j=i^{\prime}}=0$ and $\phi_{j=N-i^{\prime}}=\pi$. The associated subset of eigenenergies obtained from Eq. (27) is given by $\varepsilon^{(+)}_{\infty\pm}(p)=\beta_{1}\pm v_{0}p$ and $\varepsilon^{(-)}_{\infty\pm}(p)=-\beta_{1}\pm v_{0}p$. These limits are just the four cone-like branches of the envelope described in Eq. (22), delimiting the domain of existing eigenenergies of $H^{(N)}_{\mathrm{bulk}}(\varepsilon)$. The bulk subbands inside the envelope form a continuum as $N$ approaches infinity. The continuum is the co-subset associated with $\phi_{j}\neq0$. Furthermore, among others the two branches $\varepsilon^{(+)}_{\infty-}(p)$ and $\varepsilon^{(-)}_{\infty+}(p)$ touch each other at $(p=p_{D}=\beta_{1}/v_{0},\varepsilon=0)$, where the limit of the edge momentum $p_{\mathrm{edge}}^{(\infty)}$ is reached and the bulk gap closes up with $\Delta_{\mathrm{gap}}^{(\infty)}=0$ according to Eq.s (51) and (52). By recombining segments of $\varepsilon^{(+)}_{\infty-}(p)$ and $\varepsilon^{(-)}_{\infty+}(p)$ and redefining energy $\varepsilon^{(\pm)}_{\infty}(p)$ with respect to $(p=p_{D},\varepsilon=0)$, it is easy to obtain
\begin{equation}
\varepsilon^{(\pm)}_{\infty}(p)=\pm v_{0}|p-p_{D}|.
\end{equation}
Therefore, the envelope of the bulk domain is a linear annular cone apexed along the projection of the DP spiral. Correspondingly, the crossover of the eigenmodes can be understood by replacing energy $\varepsilon$ in Eq. (44) by each of the branches of the envelope. As a result, the infinite limit $\mathcal{\underline{U}}_{\infty}$ of the normalized eigenmodes associated with $\varepsilon^{(\pm)}_{\infty}(p)$ (the $\pm$ sign in the subscript does not matter and is omitted) is given by components
\begin{equation}
\left(
\begin{array}{c}
\mathcal{U}_{\infty,B_{l+1}}\\
\mathcal{U}_{\infty,A_{l}}
\end{array}%
\right)
=\frac{1}{\sqrt{2}}\left(
\begin{array}{c}
\pm1\\
1
\end{array}%
\right),
\end{equation}
with arbitrary $l=1,2,\ldots$.

The elucidation concludes with the calculation results in the bulk limit. Using Eq. (27), the bulk subbands in ABCNG can be calculated for a huge number ($N$) of layers as desired. In Fig. 6(a), the results for $N=1000$ are plotted, showing a practically gapless domain full of bulk subbands. The envelope is describable by the linear annular Dirac cone described in Eq. (53). The bulk DOS for the same $N$ is calculated from Eq. (49) and plotted in Fig. 6(b), where those features present for finite $N$ (Fig. 5) are practically completely smeared. In agreement with previous researches,\cite{guinea06} the calculated bulk DOS, being linear in energy $\varepsilon$ and vanishing at $\varepsilon=0$, is of the same form as in monolayer graphene.\cite{rammal85} The characterization by the bulk DOS reflects the fact that RG is a semimetal with half-filled DP spirals of the $3$D Dirac cone structure in the bulk.

\par\noindent
\section{CONCLUSION}
The physics of layered systems in ABC-stacking configuration is interesting. Compared to ABC-stacked $3$D topological insulators such as Bi$_{2}$Se$_{3}$ and Bi$_{2}$Te$_{3}$, RG behaves as a topological semimetal. It possesses flat surface subbands while being semimetallic in the bulk with the $3$D Dirac cone structure, whose DPs form spiraling lines in momentum space. There is a bulk-surface correspondence due to the non-trivial topology induced in this configuration. We studied the evolution of the gapped bulk subbands in ABCNG with increasing the number ($N$) of graphene layers under the effect of coupling to the surface subbands. The bulk gap was shown to close up at the DP spirals in the bulk limit. We elucidated the dimensional crossover of the bulk subbands to the $3$D Dirac cone structure. The coupling effect on the bulk subbands was clarified by means of the non-perturbative effective bulk Hamiltonian based the finite chain model. It was shown that as a consequence the wave length of the standing wave function across the stack of layers (along the $z$ direction) depends on the in-plane Bloch momentum. The coupling vanishes in the bulk limit and hence, the $k_{z}$ wave number is irrelevant to $p$.

The minimal model we used suffices for the purpose of the present work. Inclusion of extra hopping integrals would cause high-order anisotropy but dose not change the evolution of the bulk subbands. The major influence of on-site Coulomb potential is the splitting of the surface subbands, which can be experimentally conditioned to be zero for realizing the $2$D nature of $3$D QHE, not to mention its diminishing and vanishing role for large $N$. We remark that this splitting of the surface subbands is probably an alternative clue to elucidating the dimensional crossover. As observed in ABC-layered topological insulators, Bi$_{2}$Se$_{3}$,\cite{zhangy10} dimensional crossover occurs with reducing the number of layers, where the surface subbands are gapped in the $2$D limit. Nevertheless, in ABC-stacked graphene the gapping of surface subbands is caused by on-site Coulomb potential, whereas for ABC-layered topological insulators the gapping arises from the coupling between surface states in the two outmost layers.\cite{zhangy10} It is well known that a topological insulator behaves as an insulator in its bulk while it has conductive surface subbands described by a Dirac cone. In essence, RG and ABC-layered topological insulators both contain $2$D Dirac fermions in the bulk and the surfaces, respectively. As such, their dimensional crossovers are strikingly the same in that both of their Dirac cones become gapped subbands in the $2$D limit.

\begin{acknowledgments}
This work was supported by the Ministry of Science and Technology of Taiwan,
under the Grant nos. MOST 103-2811-M-165-001 and NSC
102-2112-M-165-001-MY3.
\end{acknowledgments}

\newpage

\bigskip \vskip0.6 truecm\newpage
\centerline {\Large \textbf {Figure
Captions}}
\begin{itemize}

\item[FIG. 1.](Color online) (a) Stacking configuration of ABCNG, with layers labelled by numerals. $B_{1}$ (filled), $A_{N}$ (unfilled): surface sublattices, shown large with orange circles; $B_{l\neq1}$ (filled), $A_{l\neq N}$ (unfilled): bulk sublattices. The TB hoppings $\beta_{0}$ and $\beta_{1}$ are shown between representative atoms. One representative chain is shown by linked thick sticks. (b) Solid view of the lattice of ABCNG and alternatively, a view of the bulk lattice of RG by infinitely extending the number of layers. The primitive unit vectors $\mathbf{a}_{1,2,3}$ of the bulk of RG add up to the $c$-axis, where the primitive unit cell and alternative hexagonal unit cell are, respectively, shown by the rhombohedron (red) and hexahedron (blue).

\item[FIG. 2.](Color online) (a) Rhombohedral BZ (red) accompanied with a prism (blue) belonging to the folded hexagonal zone. Filled (blue) and unfilled dots are high-symmetry points of the folded hexagonal zone, only the unfilled ones coinciding high-symmetry points of the rhombohedral BZ. The $2$D BZ (black) of the projected $c$-axis plane is plotted above, with $\bar{K}^{(\pm)}$ (filled in blue) being the projections of edge lines of the folded hexagonal zone. (b) A vertical cut of (a), taken through suitable coincident high-symmetry points. The folded hexagonal zone is achieved by folding wedge $z_{2}$ to $z_{1}$ and $z_{3}$ to $z_{4}$. The solid line (black) is laid at $k_{z}=0$, being equivalent to the $2$D BZ shown above.

\item[FIG. 3.](Color online) (a) Projections (red) of the DP spirals on the $2$D BZ, where the portions inside and outside are plotted in solid and dotted arcs, respectively. The arrows indicate the spiraling senses in the increase of $k_{z}$. (b) Scale-up of the projection around the specified $\bar{K}^{(+)}$ in (a). Twelve DPs are marked, for which the numerals denote the associated values of $k_{z}$ in the unit of $\pi/(6d)$ (mod $2\pi/d$). The clockwise spiraling sense is indicated by the arrow.

\item[FIG. 4.](Color online) Band structure of ABCNG. Flat pair about the zero energy (grey): surface subbands; the rest (red): bulk subbands. The direction of in-plane momentum $p$ is arbitrary.

\item[FIG. 5.](Color online) Bulk DOSs in ABCNG, in the unit of number of states per $\beta_{1}$ per atom.

\item[FIG. 6.](Color online) (a) Bulk subbands in ABCNG calculated with $N=1000$. The direction of in-plane momentum $p$ is arbitrary. (b) Bulk DOS associated with (a), in the unit of number of states per $\beta_{1}$ per atom.

\end{itemize}

\begin{figure}[p]
\includegraphics[scale=0.8]{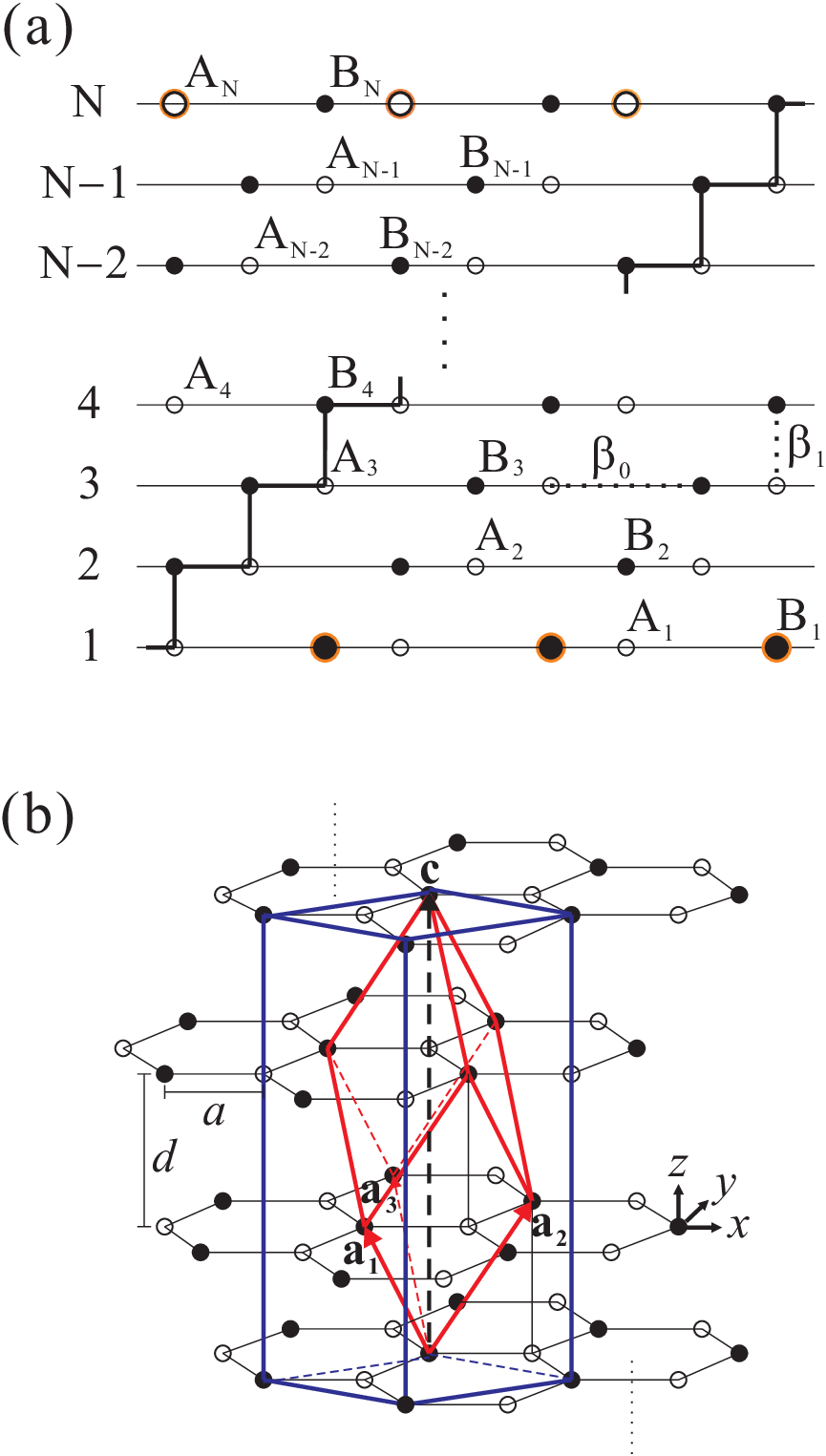}
\end{figure}

\begin{figure}[p]
\includegraphics[scale=0.8]{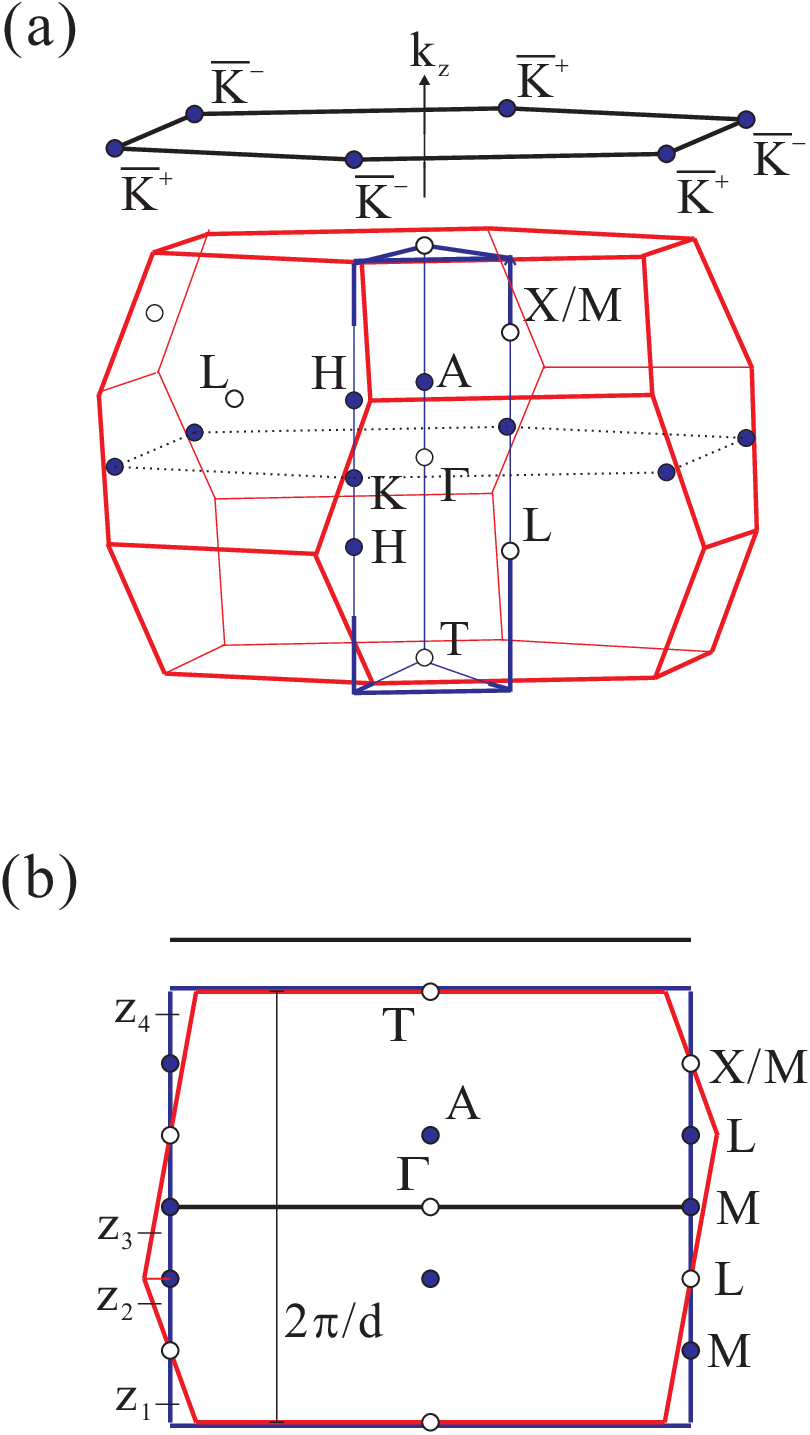}
\end{figure}

\begin{figure}[p]
\includegraphics[scale=0.8]{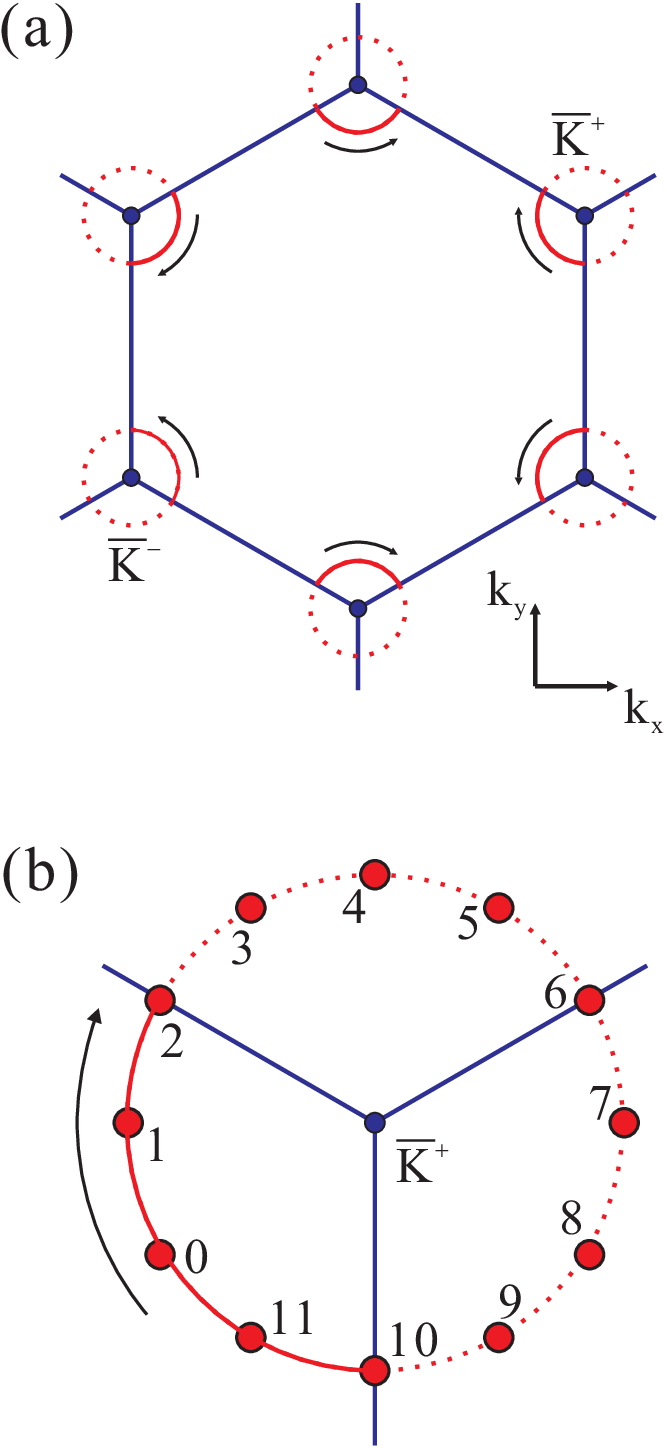}
\end{figure}

\begin{figure}[p]
\includegraphics[scale=0.8]{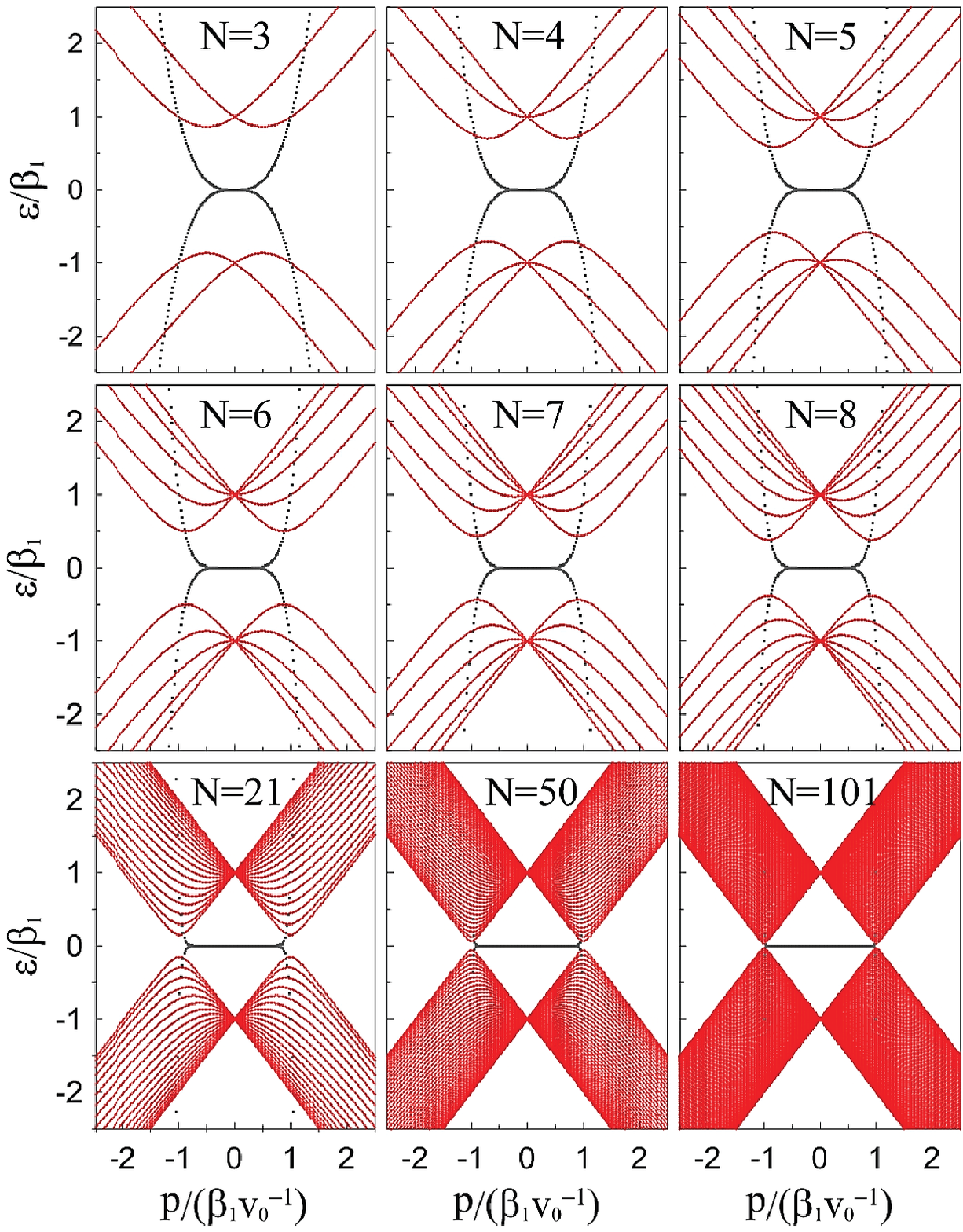}
\end{figure}

\begin{figure}[p]
\includegraphics[scale=0.8]{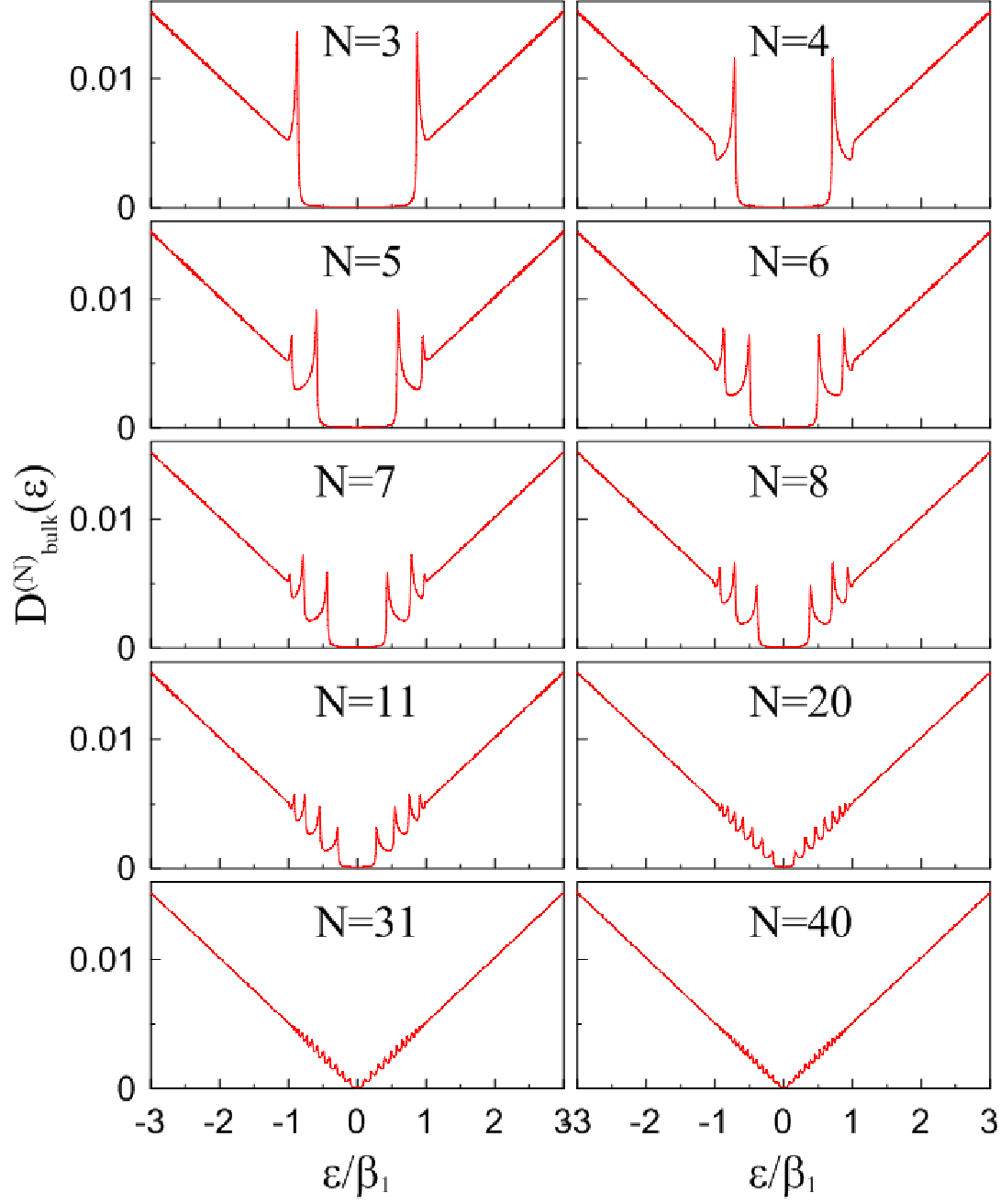}
\end{figure}

\begin{figure}[p]
\includegraphics[scale=0.8]{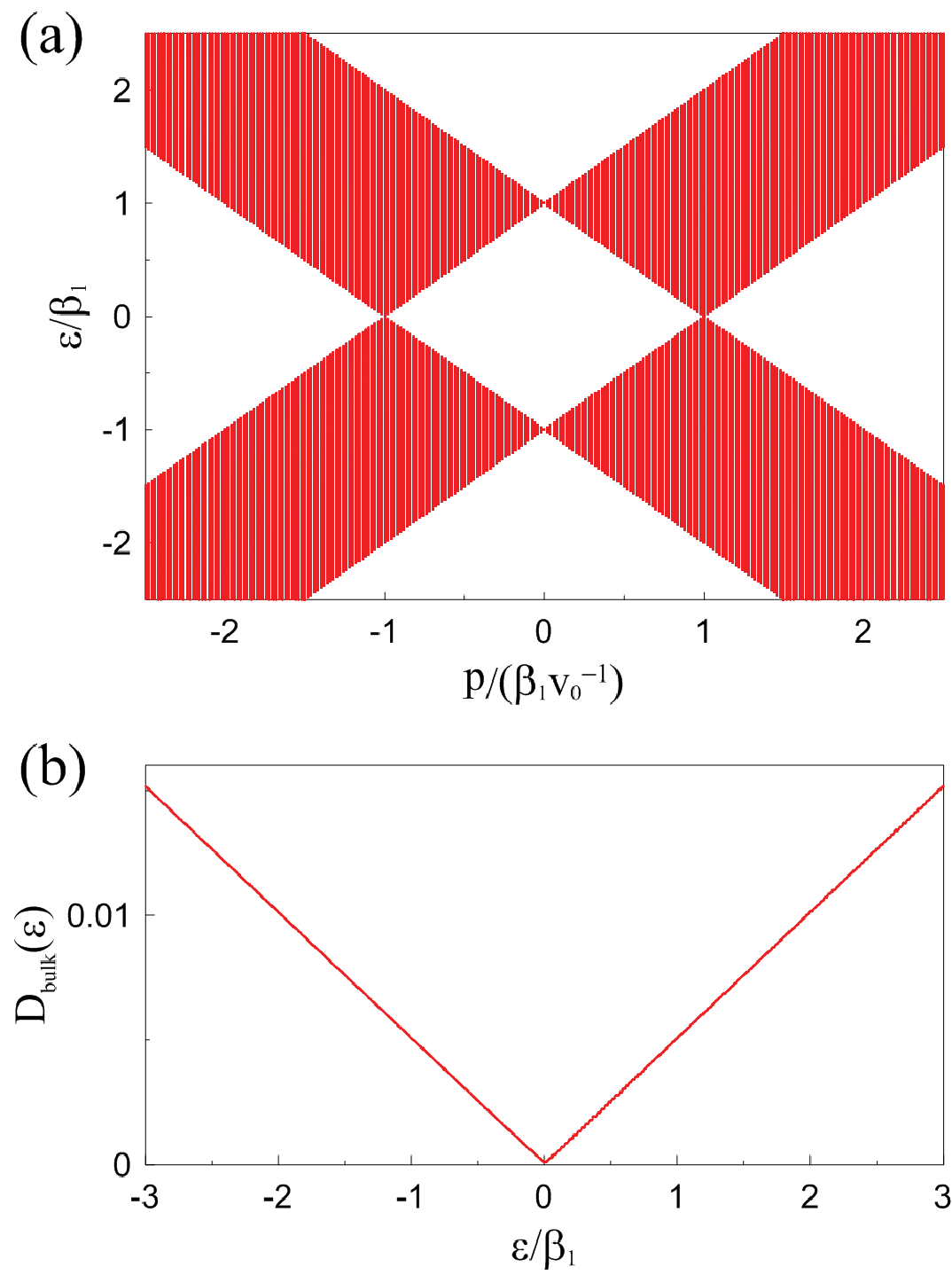}
\end{figure}

\end{document}